\documentclass[aps,prl,preprint,superscriptaddress,showpacs,floats]{revtex4-2}
\usepackage{graphicx}
\usepackage{amsfonts}
\usepackage{amsmath}
\usepackage{amssymb}
\usepackage{exscale}
\usepackage{afterpage}				
\usepackage{lineno}
\usepackage{diagbox}
\usepackage{xcolor}
\usepackage{xfrac}
\begin{document}

\title{Growth orientation and magnetic properties of GdVO$_3$ tailored by epitaxial strain engineering}

\author{M. Martirosyan}
\email{mariammartirosyan@hotmail.fr}
\affiliation{Institut Jean Lamour, CNRS, Universit\'e de Lorraine, F-54000 Nancy, France}

\author{A. Gudima}
\affiliation{Department of physics and materials science, University of Luxembourg, 41 rue du Brill, L-4422 Belvaux, Luxembourg}

\author{J. Varignon}
\affiliation{CRISMAT, CNRS, Normandie Universit\'e, ENSICAEN, UNICAEN, F-14000 Caen, France}

\author{J. Ghanbaja}
\affiliation{Institut Jean Lamour, CNRS, Universit\'e de Lorraine, F-54000 Nancy, France}

\author{S. Migot}
\affiliation{Institut Jean Lamour, CNRS, Universit\'e de Lorraine, F-54000 Nancy, France}

\author{A. David}
\affiliation{CRISMAT, CNRS, Normandie Universit\'e, ENSICAEN, UNICAEN, F-14000 Caen, France}

\author{M. Guennou}
\affiliation{Department of physics and materials science, University of Luxembourg, 41 rue du Brill, L-4422 Belvaux, Luxembourg}

\author{O. Copie}
\email{olivier.copie@univ-lorraine.fr}
\affiliation{Institut Jean Lamour, CNRS, Universit\'e de Lorraine, F-54000 Nancy, France}


\begin{abstract}
Transition-metal oxides with an ABO$_3$ perovskite structure exhibit significant coupling between spin, orbital, and lattice degrees of freedom, highlighting the crucial role of lattice distortion in tuning the electronic and magnetic properties of these systems. In this study, we grow a series of antiferromagnetic GdVO$_3$ thin films on different substrates introducing various strain values. The results demonstrate that the strain not only affects the material's structure and magnetic transition temperature but also influences the growth direction of the orthorhombic unit cell. A correlation is observed between the orientation of the orthorhombic long axis direction and the material's Néel temperature. DFT calculations highlighting the link between strain, lattice distortions and magnetic characteristics confirm these conclusions. Our findings demonstrate that the development of novel features in RVO$_3$ through material design requires control of growth orientation through strain engineering.
\end{abstract}

\maketitle
\newpage
%

\section{Introduction}
Atomic displacements in transition metal oxides (TMO) with a perovskite structure (ABO$_3$) are fundamental to the interplay between the charge, orbital, and spin degrees of freedom of the electrons at the B-site~\cite{Imada1998,Tokura2000,Zubko2011}. Structural distortions modify the electronic properties and transfer integral between neighboring atoms, leading to complex orbital and spin order phase diagrams in TMO~\cite{Goodenough2004,Dagotto2005}. Beyond the ideal, undistorted cubic structure, orthorhombic structures are commonly observed in ABO$_3$ compounds. This structural adaptation arises from the ability of the lattice to accommodate mismatches between A-O and B-O bond lengths, a characteristic also exhibited in the RVO$_3$ series. 

RVO$_3$ (R is a rare-earth or yttrium) compounds have garnered significant interest due to their potential for tailoring $t_{2g}$ orbital physics~\cite{Radhakrishnan2021,Meley2021}, resulting in intriguing magnetic~\cite{Ren1998}, optical~\cite{Wang2015,Zhang2017}, and electronic properties~\cite{Hotta2007,Zhou2007}. Cooperative rotations of VO$_6$ octahedra result in an orthorhombic $Pbnm$ structure with an $a^-a^-c^+$ tilt pattern according to Glazer's notation~\cite{Glazer1972} at room temperature~\cite{Miyasaka2003,Sage2007}. However, oxygen octahedra are not rigid and atomic displacements extend beyond those induced by antiferrodistortive (AFD) motions. This is primarily characterized by long and short V$-$O bond lengths along the [100] and [010] pseudocubic directions~\cite{Zhou2005}. The Jahn-Teller (JT) distortion of the octahedra causes these bond length variations, lifting the degeneracy between the JT-active V $t_{2g}$ levels, ideally resulting in two electrons per V$^{3+}$ site occupying either the $d_{xy}$ and $d_{xz}$ orbitals or the $d_{xy}$ and $d_{yz}$ orbitals. Below room temperature, the JT distortions of the two $t_{2g}$ electrons at V$^{3+}$ sites subtly influence the physical properties~\cite{Zhou2008} and promote a G-type ordering (G$_{\rm{OO}}$) of the occupied orbitals at the temperature T$_{\rm{OO}}$, revealed at the macroscopic level by a structural phase transition from orthorhombic $Pbnm$ to monoclinic $P2_1/b$~\cite{Sage2007,Miyasaka2003}. At lower temperatures, superexchange interactions promote a C-type antiferromagnetic spin order (C$_{\rm{SO}}$) of the V spins at the Néel temperature T$_{\rm{N}}$. As the R$^{3+}$ ionic size decreases, the cooperative octahedral rotations increase, leading to a reduced overlap integral of the V $3d$ and O $2p$ orbitals. This, in turn, causes T$_{\rm{N}}$ to decrease monotonically with the size of R$^{3+}$. 

An approach for controlling electronic and magnetic properties in bulk RVO$_3$ relies on hydrostatic pressure or temperature \cite{Zhou2007,Zhou2009}. Instead, when grown in thin film, RVO$_3$ can be engineered through the application of biaxial strain. Firstly, epitaxial LaVO$_3$ thin films have been grown using pulsed laser deposition (PLD)~\cite{Hotta2006} or molecular beam epitaxy (MBE)~\cite{Zhang2015}. Noteworthy, the research has explored how strain affects the rotation of VO$_6$ octahedra~\cite{Rotella2012,Rotella2015} and the movement of La atoms in LaVO$_3$ thin films~\cite{Masset2020, Meley2018}. Moreover, tunable magnetic properties have been reported in PrVO$_3$ thin films, either as a function of epitaxial strain~\cite{Kumar2019a,Kumar2019b} or chemical strain through the control of the concentration of oxygen vacancies~\cite{Copie2013,Copie2017}. Additionally, RVO$_3$ compounds play an important role, as a prototypical system, for tailoring and understanding hybrid improper ferroelectricity~\cite{Bousquet2008,Benedek2011,Rondinelli2012} and magnetoelectricity \textit{by design} as proposed in symmetry-breaking RVO$_3$ based superlattices~\cite{Varignon2015}.

Among the RVO$_3$ series, GdVO$_3$ combines several characteristics, which make this compound interesting. Gd$^{3+}$ has a $4f^7$ electronic configuration and \textit{a priori} carries no orbital moment. At room temperature, bulk GVO adopts an orthorhombic $Pbnm$ crystal structure, with lattice parameters: $a_o = 5.339$~\AA, $b_o = 5.589$~\AA, and $c_o = 7.632$~\AA~($o$ stands for orthorhombic)~\cite{Sage2007}. Considering a pseudocubic ($pc$) unit cell approximation, $a_{pc}\approx a_o/\sqrt2\approx b_o/\sqrt2\approx c_o/2$, the averaged subcell parameter $a_{pc}$ is 3.847~\AA. GVO undergoes a structural phase transition to the monoclinic space group $P2_{1}/b$, along with an orbital ordering at T$_\text{OO}$ = 199 K, followed by a C$_{\rm{SO}}$ ordering at T$_\text{N}$ = 118 K \cite{Miyasaka2003,Tung2006}. Below this transition temperature, Sage \textit{et al.} demonstrate a coexistence region of C-type (C$_{\rm{OO}}$) and G-type (G$_{\rm{OO}}$) orbital phases in compounds with intermediate rare earth elements such as Gd \cite{Sage2007}. These orbital orders are linked to the spin orders via the Goodenough-Kanamori rules, with C$_{\rm{OO}}$ coinciding with G$_{\rm{SO}}$, while G$_{\rm{OO}}$ is associated with C$_{\rm{SO}}$ \cite{Goodenough1955, Goodenough1958, Kanamori1959}. Additionally, a low-temperature phase transition occurs at T$_\text{M}$ = 8 K, characterized by the emergence of an unusual magnetic memory effect \cite{Tung2006}. The coexistence of several orbital and spin order phases makes this compound particularly interesting for studying the relationship between structure, properties and strain engineering of spin-orbital interactions.

Here, we report on the epitaxial growth of GdVO$_3$ (GVO) thin films by pulsed laser deposition (PLD) on various substrates. We investigate the effect of the substrate-induced biaxial strain on the structural and magnetic properties of the films. The investigation of the structure and the microstructure unveil the existence of different orientations of the $[001]_o$ direction in the epitaxial GVO thin films. Strain engineering enables the tuning of the antiferromagnetic transition temperature and a link is observed between the orientation of the $[001]_o$ direction and this temperature. First principles calculations reveal that epitaxial strain engineering has a strong impact on the structure and the magnetic ground state. This approach allows thus tailoring the lattice distortions and highlights the subtle yet complex growth mechanisms that can lead to coexistence of structural, magnetic, and electronic phases in this compounds.
\section{Methods}

\textit{Experimental details}. GVO thin films were synthesized using PLD. A KrF excimer laser ($\lambda = 248$~nm) was focused on a GdVO$_4$ ceramic target at a frequency of 1 Hz and a fluence of 2~J/cm$^2$. In this study, we varied the biaxial strain by using (001)-oriented substrates such as LaAlO$_3$ (LAO), SrLaGaO$_4$ (SLGO), (La$_{0.18}$Sr$_{0.82}$)(Al$_{0.59}$Ta$_{0.41}$)O$_3$ (LSAT), SrTiO$_3$ (STO), and KTaO$_3$ (KTO), with a cubic or pseudocubic lattice parameters of 3.790~\AA, 3.843~\AA, 3.868~\AA, 3.905~\AA, and 3.989~\AA, respectively. For simplicity, both cubic and pseudo-cubic descriptions are labeled as pc. Before the growth, the substrate surface was cleaned by using an oxygen plasma with a RF power of 300 W at an oxygen pressure of $5\times5.10^{-5}$ mbar for 5 minutes (except for LSAT and KTO). All films were deposited at an optimized growth pressure and temperature of $10^{-7}$ mbar and 900$^{\circ}$C, respectively. The film growth was monitored with \textit{in situ} reflection high-energy electron diffraction (RHEED) and real-time monitoring of the RHEED intensity oscillation was used for a precise control over the film thickness, which is around 30~nm. A capping layer consisting of 4-5 unit cells of LAO was also grown at the end of the growth to protect the surface from over-oxidation~\cite{Kumar2019b}. The surface structure was characterized using a Park NX10 atomic force microscope (AFM), and the crystalline structure was investigated by X-ray diffraction (XRD) using a PANalytical X’pert Pro MRD diffractometer with monochromatic Cu $K_{\alpha_{1}}$ radiation ($\lambda = 1.54056$ \AA). Additionally, the thickness was also determined by X-ray reflectivity (XRR). Raman scattering measurements were performed in the backscattering configuration using a Renishaw InVia Raman spectrometer in micro-Raman mode with a 100X objective. The spectra were collected at room temperature using 633 nm He-Ne laser for the excitation. The atomically resolved microstructure was observed by high-resolution transmission electron microscopy (HRTEM) and high-angle annular dark field (HAADF) scanning transmission electron microscopy (STEM). These observations were conducted using a double-aberration corrected cold FEG JEOL ARM200F microscope operating at 200 kV, and equipped with Energy-dispersive X-ray spectroscopy (EDS). Prior to the observations, the cross-section were prepared by using a focused ion beam (FIB) with Ga$^{+}$ ion milling (FEI HELIOS NanoLab 600i). Magnetic measurements were conducted using a Quantum Design superconducting quantum interference device magnetometer (MPMS3). 

\textit{Theoretical details}. First principles calculations were performed using density functional theory (DFT) with the VASP package~\cite{Kresse1993,Kresse1996} and the PBEsol functional for modeling exchange correlation phenomena. In addition, we considered a U potential on V $3d$ levels of 3.5 eV, entering as a single effective parameter~\cite{Duradev1998}. This parameter was fitted providing correct electronic, magnetic, and structural features for the GVO ground state \cite{Varignon2015}. Gd $4f$ electrons are not explicitly considered here and are included in the projected augmented wave (PAW) potential~\cite{Blochl1994} to amend self interaction errors inherent to the implementation of DFT. In the calculations, a ($2a$, $2a$, $2a$) cubic cell was considered, which allows for the oxygen cage rotations and Jahn-Teller motions to develop. The energy cutoff is set to 500 eV and a $4\times4\times4$ $k$-point mesh is employed. We have considered two growth orientation, \textit{i.e} $[001]_o$ and $[110]_o$, and two V spin orderings, \textit{i.e} G$_{\rm{SO}}$ and C$_{\rm{SO}}$. We then block two GVO lattice parameters to those of the substrate and relax the magnitude of the remaining lattice parameter. As evidenced in PrVO$_3$ thin films, the presence of structural domains or variants prevent the crystallographic axis along the growth direction from tilting~\cite{Copie2017}. We have therefore performed the calculations either restricting the axis to be orthogonal to the substrate or keeping it free to tilt.

\section{Results and discussions}
%
%
%
%
%
\subsection{Growth and structure of GVO thin films on cubic substrates}
\begin{figure*}[tp]
    \begin{center}
        \includegraphics[ clip,width=1\textwidth]{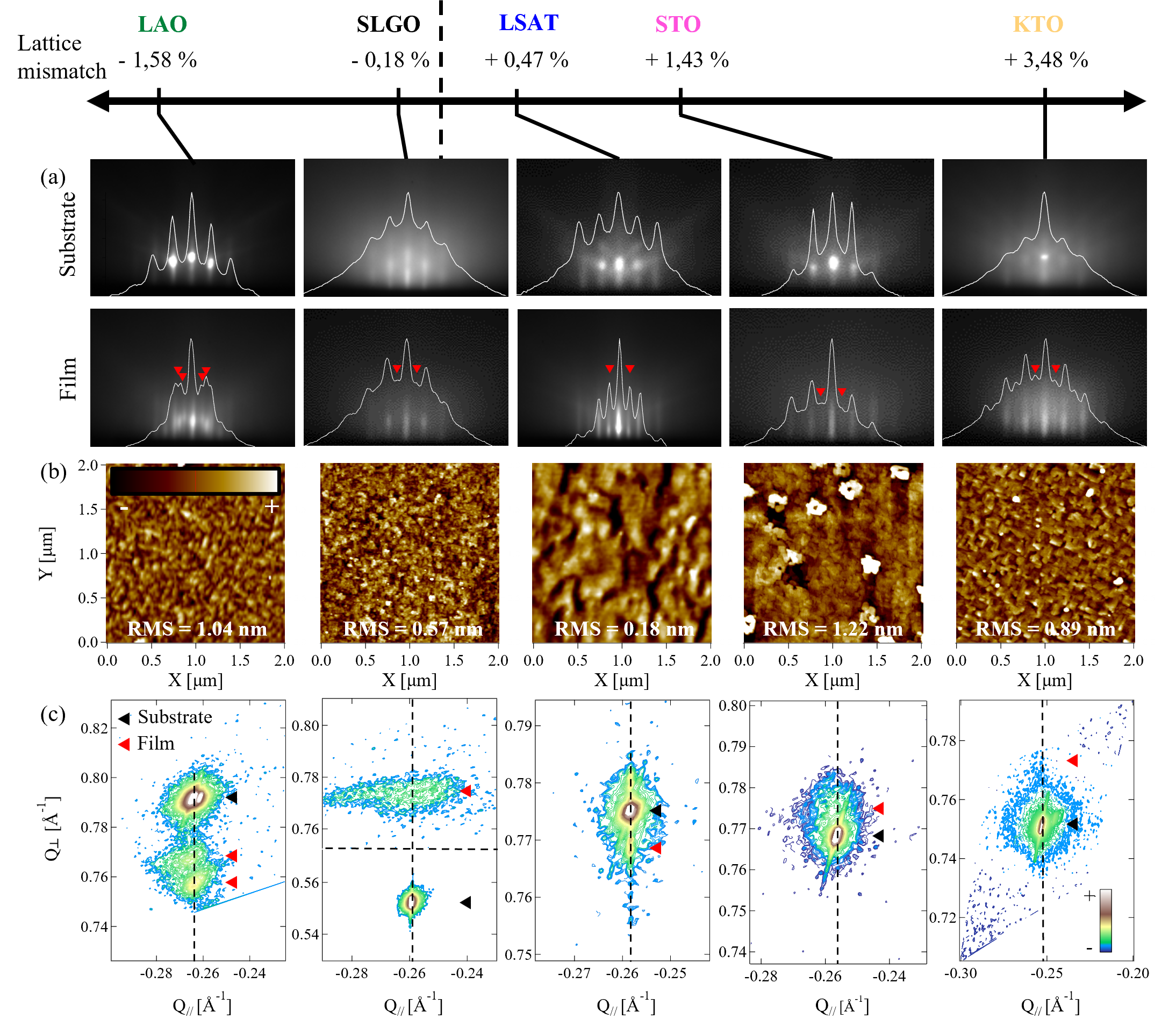}
    \end{center}
\caption{(a) Top: RHEED patterns of substrates LAO, SLGO, LSAT, STO, and KTO in the $[100]$ direction arranged by strain from left to right. Bottom: RHEED patterns after the growth of around 30 nm thick GVO films. (b) AFM images ($2~\mu m\times 2~\mu m$) of the film surfaces after capping with 4-5 unit cells of LAO. (c) Reciprocal space map around the $(103)_{pc}$ pseudocubic reflection, with the substrate peak indicated by a black arrow and the film peak by a red arrow. The alignment of the peaks from the film and substrate suggests a fully strained film.}
\label{Fig1} 
\end{figure*}

We have tuned the biaxial strain by growing a series of GVO thin films on various substrates ranging from a compressive strain state for LAO (-1.58$\%$) and SLGO (-0.18$\%$) to a tensile strain state, for LSAT (+0.47$\%$), STO (+1.43$\%$), and KTO (+3.48$\%$). In Fig.~\ref{Fig1}(a), we show the RHEED patterns of the substrates and films in the $[100]_{pc}$ pseudocubic direction. The main streaks of the GVO pattern match the substrates ones and indicate the epitaxial growth on (001)-oriented substrates. We have included white line profiles for each pattern to further illustrate the correspondence of the streaks. In addition to the main peaks, small peaks (indicated by red arrows) are also observed. These half-order extra reflections are not related to a simple perovskite lattice and correspond to a doubling of the unit cell along the $[100]_{pc}$ direction (also seen in the $[010]_{pc}$ direction, not shown). The presence of these superstructure agrees with $[001]_o$ lying in the substrate $[100]_{pc}$ or $[010]_{pc}$ directions. During the deposition on LAO, LSAT and STO, we observed the oscillations of the RHEED intensity (not shown) that indicates a GVO layer-by-layer growth mode.
Figure~\ref{Fig1}(b) displays the corresponding AFM images of the grown films. Despite the presence of a LAO capping layer, the surface roughness we determined is a fair indication of the GVO surface roughness since the LAO layer adopts a layer-by-layer growth on LSAT, LAO, SLGO and STO. The surface roughness is rather low for the film grown on LSAT and it is 2-3 units cells for the others.
In Fig.~\ref{Fig1}(c), we show the reciprocal space maps (RSM) around the film and substrate asymmetric $(103)_{pc}$ Bragg peaks, indicated by a red and a black arrow, respectively. The mapping reveals well-defined peaks of the film for LAO, SLGO, LSAT and STO. On the contrary, the GVO Bragg peak has a very low intensity for KTO and seems not aligned with the substrate one, indicating a film relaxation for such a high expected strain value (+ 3.48$\%$). For the other substrates, the vertical alignment of the substrate and film peaks shows that the film are fully strained. For tensile strain (LSAT and STO), the film's peak width is small in the Q$_{||}$ direction. On the contrary, several peaks are observed for compressive strain (LAO and SLGO). Two peaks appear for GVO, separated in the Q$_{||}$ and in the Q$_{\perp}$ directions for SLGO and LAO, respectively. This suggests either the coexistence of a strained and relaxed phases or the coexistence of polymorphic structural phases. In the latter case, the coexistence of twins or variants of GVO, whose $[110]_o$ direction could tilt with respect to the $[001]_{pc}$ direction, similar to a monoclinic distortion of the pseudocubic unit cell, may occur~\cite{Ko2011}. 

%
%
%

\begin{figure*}[tp]
    \begin{center}
        \includegraphics[width=1\textwidth]{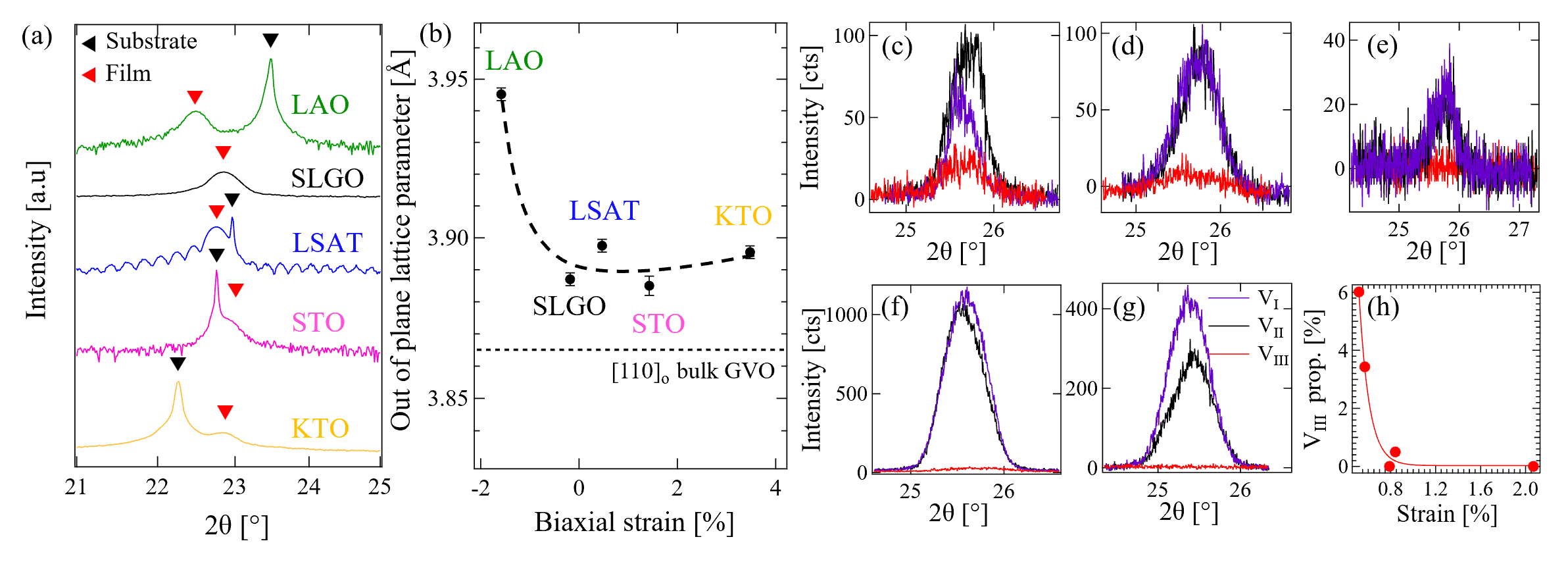}
    \end{center}
\caption{ (a) $2\theta-\theta$ XRD measurements around the pseudocubic $(001)_{pc}$ reflections of GVO films. (b) Out of plane lattice parameters extracted by a fit of the $(001)_{pc}$ reflections of GVO films as a function of lattice mismatch (the dotted line serves as a visual guide). (c)-(g) $2\theta$-scans through the $(0~\frac{1}{2}~1)_{pc}$, $(\frac{1}{2}~0~1)_{pc}$, $(1~0~\frac{1}{2})_{pc}$ Bragg peaks for GVO grown on STO, SLGO, KTO, LSAT and LAO, respectively, arranged from the lowest to the highest strain along the growth direction. These scans highlight the film growth orientations and the presence of different structural orientations: V$_{\rm I}$ corresponding to $(\frac{1}{2}~0~1)_{pc}$, V$_{\rm II}$ to $(0~\frac{1}{2}~1)_{pc}$, and V$_{\rm III}$ to $(1~0~\frac{1}{2})_{pc}$. (h) Proportion of V$_{\rm III}$ domains as a function of the strain along the growth direction.}
\label{Fig2} 
\end{figure*}

Figure~\ref{Fig2} shows the results of XRD measurements. In Fig.~\ref{Fig2}(a), high-resolution $\theta-2\theta$ measurements obtained around the $(001)_{pc}$ reflections of the substrates (indicated by a black arrow) and the $(001)_{pc}$ reflection of GVO films (indicated by a red arrow) are displayed. The presence of Laue fringes for the film grown on LSAT, but not for the other substrates, is in agreement with the AFM measurements and confirms smoother surface and interface for GVO grown on LSAT, compared to the rougher films grown on the other substrates.
In Fig.~\ref{Fig2}(b), we show the out-of-plane lattice parameter plotted against the lattice mismatch. The out-of-plane lattice parameter increases for compressive biaxial strain and decreases for tensile biaxial strain. Overall we note that the out-of-plane lattice parameter is above the bulk value for tensile strain (even for the film on KTO, which is assumed to be relaxed). This is likely due to the incorporation of oxygen vacancies during the film growth at quite low partial oxygen pressure, whose presence results in an additional elongation along the growth direction~\cite{Copie2017}. These findings highlight the intricate relationship between substrate choice, lattice parameters, and film characteristics, emphasizing the importance of meticulous analysis in epitaxial film studies.

Figures~\ref{Fig2}(c)-(g) show $2\theta$-scans of $(\frac{1}{2}~0~1)_{pc}$, $(0~\frac{1}{2}~1)_{pc}$ and $(1~0~\frac{1}{2})_{pc}$ Bragg reflections for all films. The first two peaks indicate the alignment of the long orthorhombic $c_o$-axis parallel to the in-plane $[100]_{pc}$ and $[010]_{pc}$ substrate direction, while the last peak is characteristic of the doubled unit cell along the growth direction, \textit{i.e.} $c_o$-axis parallel to the $[001]_{pc}$ direction. Indirectly, this reveals also the direction of GVO in-phase oxygen octahedra rotation and the Gd$^{3+}$ displacement as both are coupled with a trilinear energy term ~\cite{Varignon2015}. The Gd$^{3+}$ anti-polar motion displacement (or X$_5^-$ mode \cite{Amisi2012}) is perpendicular to $c_o$-axis. As the existence of $(\frac{1}{2}~0~1)_{pc}$, $(0~\frac{1}{2}~1)_{pc}$, $(1~0~\frac{1}{2})_{pc}$ half-order Bragg peaks is associated to different $c_o$-axis orientations, it characterizes the presence of domains or variants called V$_{\rm{I}}$, V$_{\rm{II}}$ and V$_{\rm{III}}$, respectively. In Fig.~\ref{Fig2}(c)-(g), we display the characteristic peaks for GVO thin films grown on STO, SLGO, KTO, LSAT and LAO, respectively, arranged from the lowest to the highest out-of-plane strain. We observe a predominance of V$_{\rm{I}}$ and V$_{\rm{II}}$ domains for all films, while the proportion of V$_{\rm{III}}$ decrease with the out-of-plane strain as represented in Fig.~\ref{Fig2}(h). The proportion of V$_{\rm{III}}$ is plotted after correcting the recorded intensities for volumetric and absorption variations that occur when rotating the sample~\cite{Rotella2015, Chateigner2010}. In bulk GVO, the interplanar distances are 3.865~\AA~and 3.815~\AA~in the $[110]_o$ and $[001]_o$ direction, respectively. Hence, a decrease in the measured lattice parameter along the growth direction should indicate an increase in the V$_{\rm{III}}$ proportion, which is consistent with our observations in Fig.~\ref{Fig2}(c)-(h).

\begin{figure*}[tp]
    \begin{center}
        \includegraphics[width=1\textwidth]{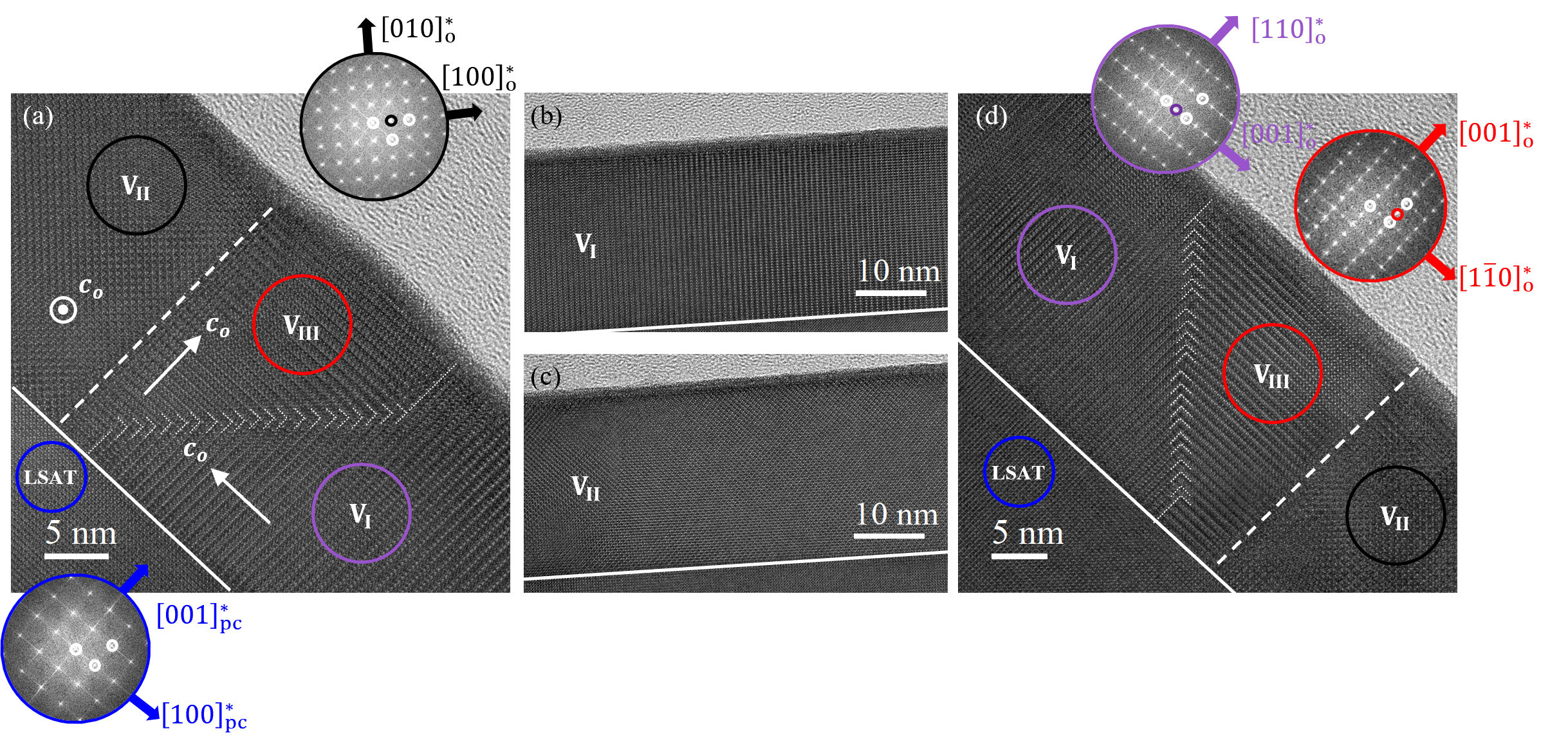}
    \end{center}
\caption{HRTEM cross-section images along the LSAT $[010]_{pc}$ zone axis taken in different regions of the lamella, showing coexistence of V$_{\rm I}$, V$_{\rm II}$ and V$_{\rm III}$ in (a) and (d), preponderance of V$_{\rm I}$ (b) or V$_{\rm II}$ (c). In the film, the three oriented domains or variants are identified according to the Fast Fourier transforms (FFT) analysis. Compared to the FFT of the cubic LSAT substrate, visible additional reflections are encircled in the orthorhombic V$_{\rm I}$, V$_{\rm II}$ and V$_{\rm III}$ domains.}
\label{Fig3} 
\end{figure*}

\begin{figure*}[tp]
    \begin{center}
        \includegraphics[width=0.62\textwidth]{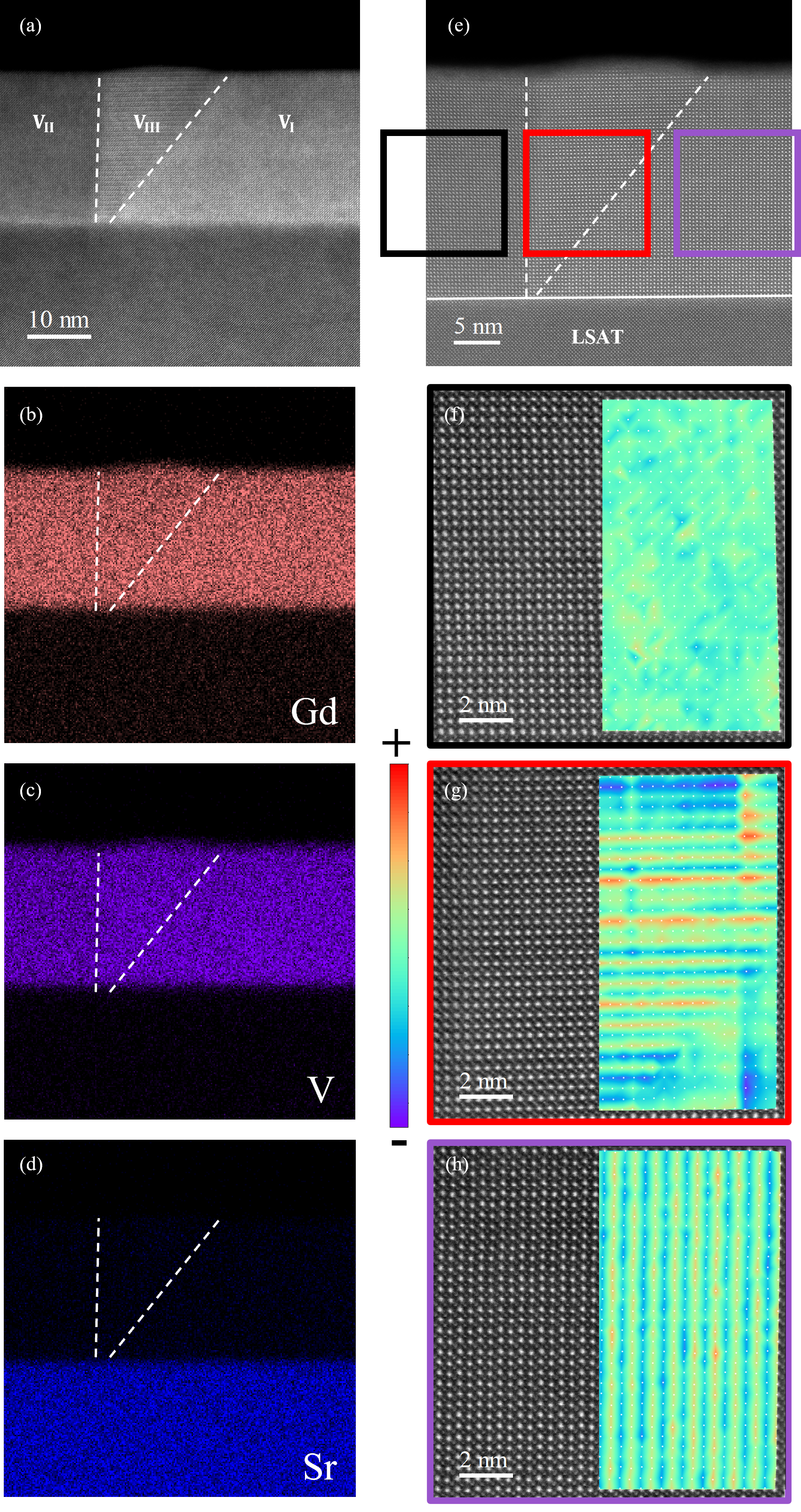}
    \end{center}
\caption{(a)-(d) EDS mapping of a GVO film grown on LSAT along the $[010]_{pc}$ zone axis. (e) HAADF STEM observations of the same film with three squared regions whose magnified images are displayed in the images (f), (g) and (h) below. The superimposed image analysis reveals the Gd$^{3+}$ displacements, which are characteristic of V$_{\rm{II}}$, V$_{\rm{III}}$ and V$_{\rm{I}}$ domains in (f), (g) and (h), respectively.}
\label{Fig4} 
\end{figure*}

%
%
%
%
%
%
%
To gain more insights, we have investigated the microstructure of the film displaying the best crystalline quality, Figure.~\ref{Fig3}(a) depicts a HRTEM cross-section image of GVO grown on LSAT taken along the $[010]$ cubic direction. First, the interface between the substrate and the film is sharp. Then, we see a contrasted image that, apart from the substrate, characterizes the presence of microstructural domains as usually observed in RVO$_3$ thin films~\cite{Rotella2012,Rotella2015,Meley2018,Masset2020} and confirmed by Fast Fourier Transform (FFT) [see below]. Each domain, separated by a white dashed line, corresponds to a distinct orientation of the orthorhombic structure on the cubic substrate, and is visible over the entire length of the cross-section lamella (i.e. along the $[100]_{pc}$ direction of LSAT). The purple (V$_{\rm{I}}$), black (V$_{\rm{II}}$) and red (V$_{\rm{III}}$) variants corresponds to the parallel alignment of the long-axis $c_o$ with the $[100]_{pc}$, $[010]_{pc}$ and $[001]_{pc}$ directions of LSAT, respectively. Low magnification HRTEM image shows that V$_{\rm{I}}$ and V$_{\rm{II}}$ are majority with large area for both, as represented in Fig.~\ref{Fig3}(b) and (c). Interestingly, we observed that a wedge-like V$_{\rm{III}}$ variant is intercalated between V$_{\rm{I}}$ and V$_{\rm{II}}$, as shown in Fig.~\ref{Fig3}(d). Indeed, the FFTs of different zones exhibit additional reflections, compared to a simple cubic lattice [see the substrate's FFT in blue], corresponding to the doubling of the orthorhombic unit cell due to the VO$_6$ octahedra rotation and Gd$^{3+}$ cation displacements. It should also be noted that the interface between the domains is not the same. Whereas, the domain boundary between V$_{\rm{II}}$ and V$_{\rm{III}}$ is sharp and parallel to the growth direction, the boundary between V$_{\rm{I}}$ and V$_{\rm{III}}$ is a chevron-like boundary [see Fig.~\ref{Fig3}(a) and (d)].
%
%
%
%
We have performed EDS analysis, which reveals that the GVO layer grown on LSAT is chemically homogeneous, without indications of atomic segregation or intermixing. The STEM image of the analyzed region and the selective elemental mappings are displayed in Fig.~\ref{Fig4}(a)-(d). To further explore the microstructure, we performed STEM-HAADF imaging, a representative image is depicted in Fig.~\ref{Fig4}(e). The film/substrate interface is sharp and the V$_{\rm{I}}$, V$_{\rm{II}}$ and V$_{\rm{III}}$ are present. We have highlighted the Gd$^{3+}$ positions and quantified each cations coordinates in the columns by using a numerical analysis of the STEM image~\cite{Nord2017,OConnell2020} for all three variants. Taken from the STEM image in Fig.~\ref{Fig4}(e), the results are superimposed to the images in Fig.~\ref{Fig4}(f), (g) and (h) for the V$_{\rm{II}}$, V$_{\rm{III}}$ and V$_{\rm{I}}$ domains, respectively. The mapping shows a contrast corresponding to the antipolar X$_5^-$ motion and further indicates the orientation of the $c_o$-axis, except for the V$_{\rm{II}}$ variant in Fig.~\ref{Fig4}(f), for which the $c_o$-axis lies along the zone axis. We determined the mean displacements corresponding to the X$_5^-$ mode, which is 22.14~$\pm$~0.42 pm for V$_{\rm{III}}$ and 27.99~$\pm$~0.35 pm for V$_{\rm{I}}$, within our experimental accuracy. These values are lower than those for the bulk material (36 pm) and differ according to the orientation of the $c_o$-axis. We point out that the image in Fig.~\ref{Fig4}(e) shows that the surface of the V$_{\rm{III}}$ variant looks like a bulge, as it can be seen in Fig.~\ref{Fig3}(a) and in EDS maps. The thickness we determined across V$_{\rm{I}}$ and V$_{\rm{II}}$ is 25.15~nm, while it is 26.20~nm across V$_{\rm{III}}$. As evidenced in PrVO$_3$ thin films, the existence of V$_{\rm{I}}$ and V$_{\rm{II}}$ variants leads to self-induced stress keeping the $[110]_o$ direction perpendicular to the substrate~\cite{Copie2017}, with sharp and straight domain boundaries across the thickness~\cite{Copie2013,Copie2017,Masset2020}. In GVO, as V$_{\rm{III}}$ is intercalated between V$_{\rm{I}}$ and V$_{\rm{II}}$, we propose that similar constraints exist, which could also be represented by the rotation of the free $[110]_o$ direction around $[1\bar{1}0]_o$ direction. In Fig.~\ref{Fig3}(a), we have represented the direction of $c_o$-axis of the domains. V$_{\rm{I}}$ converges on V$_{\rm{II}}$, both with $c_o$ lying in the substrate plane. V$_{\rm{II}}$ act then as a backstop, so that V$_{\rm{III}}$ straightens up and forms a bulge with $c_o$ perpendicular, similar to an accretionary wedge. Consequently, the thickness of V$_{\rm{III}}$ is larger despite a smaller interplanar distance along the $[001]_o$ direction. 

%
%
%
\subsection{DFT calculations and symmetry mode analysis}
\begin{figure*}[tp]
    \begin{center}
        \includegraphics[width=1\textwidth]{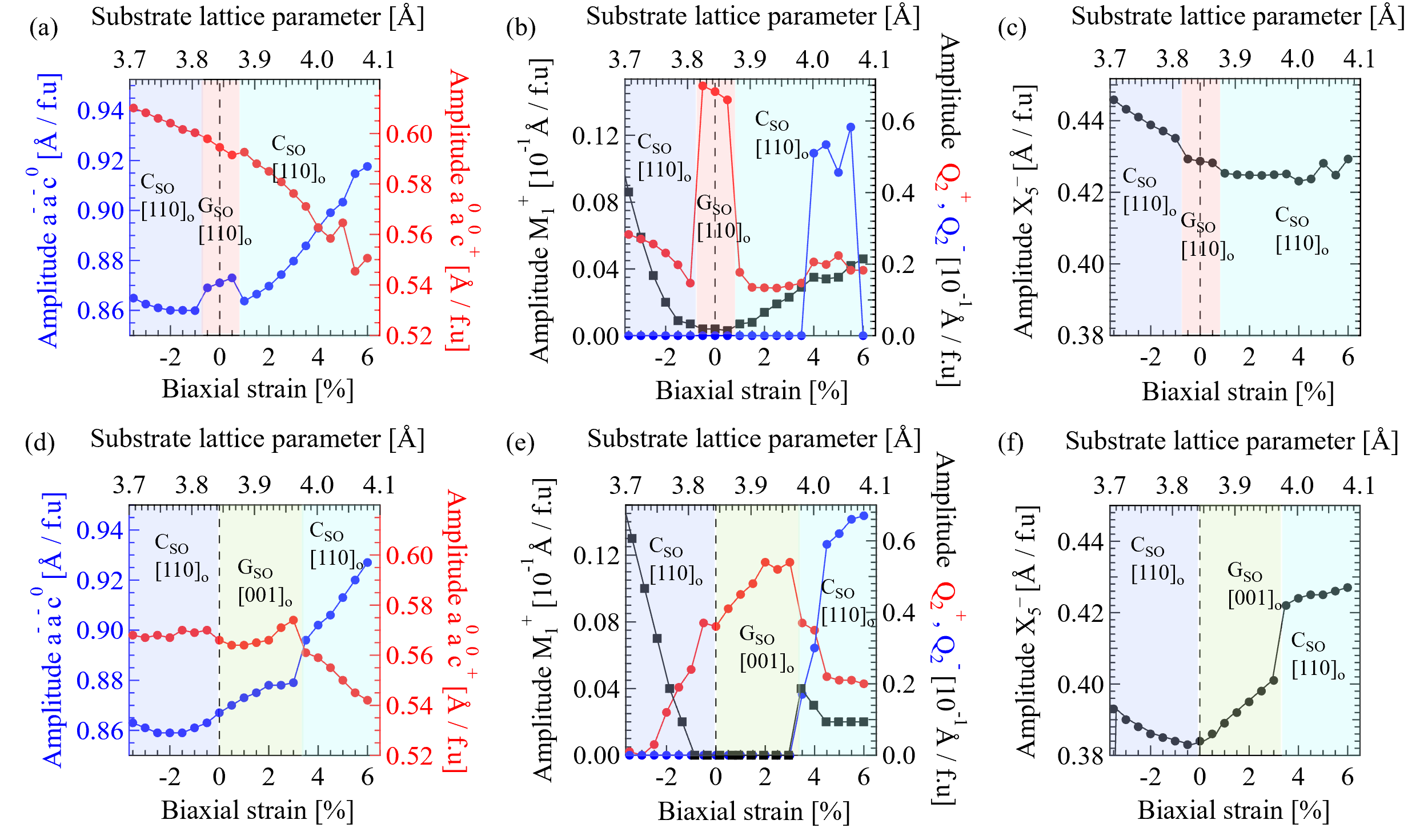}
    \end{center}
\caption{Lattice distortions and phase diagrams (magnetic ground state and growth direction) as a function of biaxial strain. (a) and (d) Oxygen octahedral rotation modes $a^-a^-c^0$ and $a^0a^0c^+$ (b) and (e) Cooperative JT modes (Q$^+_2$ in red and Q$^-_2$ in blue) and the M$^+_1$ mode in black. (c) and (f) antipolar X$^-_5$ mode. The out-of-plane direction is free to tilt in (a), (b) and (c), while it is kept orthogonal in (d), (e) and (f).}
\label{Fig5} 
\end{figure*}

To get insights on the role of the epitaxial strain on the structure and properties of GVO films, we performed first-principles simulations using DFT. We have optimized the GVO structures, (i) considering the $[001]_o$ direction being perpendicular or parallel to the substrate, (ii) assuming the out-of-plane direction to be free to tilt (presence of one variant) or restricting the out-of-plane direction to be orthogonal to the substrate (mimicking the effect of self-induced domain constraint~\cite{Copie2017}), (iii) blocking the GVO in-plane pseudocubic lattice parameters, the remaining one being free to relax, and (iv) considering either a C$_{\rm{SO}}$ or G$_{\rm{SO}}$ spin order. We have also performed a symmetry mode analysis~\cite{Orobengoa2009,Perez2010}. The derived structures are made of the oxygen octahedral rotation modes $a^-a^-c^0$ and $a^0a^0c^+$, the antipolar X$^-_5$ mode, the cooperative JT modes (Q$^+_2$ and Q$^-_2$), where the "+" sign indicates that the distortion is in phase, while the "-" sign indicates an out-of-phase distortion and the M$^+_1$ mode, which consists in the distortion of the O-V-O angles in VO$_2$ planes along the growth direction~\cite{Kumar2019a}. The results of the calculated ground states and the main distortions deduced from the geometry relaxations, as a function of the biaxial strain, are summarized in Fig.~\ref{Fig5}. The amplitudes of the octahedra rotations, JT distortions and antipolar motions for the case of the out-of-plane direction free to tilt is shown in Fig.~\ref{Fig5}(a), (b) and (c), respectively. The results considering the out-of-plane direction remaining orthogonal are shown in Fig.~\ref{Fig5}(d)-(f).

In the first case [Fig.~\ref{Fig5}(a)-(c)], the growth direction is $[110]_o$ in the considered biaxial strain range, the $[001]_o$ direction lying in the substrate plane. The magnetic ground state is C$_{\rm{SO}}$, except within a strain range of $\pm1\%$, where it becomes G$_{\rm{SO}}$. Interestingly, when the out-of-plane is forced to be orthogonal [Fig.~\ref{Fig5}(d)-(f)], the calculated ground state in the 0--3\% range corresponds to a [001]$_o$ growth orientation with a G$_{\rm{SO}}$ ordering of the V$^{3+}$ spins. Above $+$3\% and when compressively strained, GVO adopts a $[110]_o$ growth direction with a C$_{\rm{SO}}$ magnetic ground state. Thus, the role of self-induced domain constraints is very important as it allows stabilizing V$_{\rm{III}}$ with the $[001]_o$ growth direction. However, the high sensitivity of GVO properties with respect to its structure is subtle and more complex, as we observe several domains in the grown film instead of the mono-domain configuration considered in the calculations. For epitaxial GVO grown onto STO, the biaxial strain is +1.43\% and the orientation of GVO should be $[001]_o$. Experimentally, the proportion of V$_{\rm{III}}$ is indeed enhanced for the STO. However, it remains minority compared to V$_{\rm{I}}$ and V$_{\rm{II}}$. Given the larger size of V$_{\rm{I}}$ and V$_{\rm{II}}$ we observed for LSAT, it is reasonable to assume that free and fixed out-of-plane directions may coexist in the film, which in turn would lead to the coexistence of different spin and orbital orders. This could also favor the coexistence of different lattice symmetries as C$_{\rm{SO}}$ and G$_{\rm{SO}}$ are associated in the bulk material to a monoclinic $P2_1/b$ and orthorhombic $Pbnm$ symmetries, respectively.

Regarding the lattice distortions, for the out-of-plane direction free to tilt [Fig.~\ref{Fig5}(a)-(c)], biaxial strain (or substrate's choice) acts on the amplitude of the $a^-a^-c^0$ and $a^0a^0c^+$ modes. As the substrate lattice parameter increases, the $a^-a^-c^0$ amplitude increases from 0.865 to 0.928 \AA/f.u, while the $a^0a^0c^+$ one decreases from 0.610 to 0.551 \AA/f.u [see Fig.~\ref{Fig5}(a)]. Considering the JT distortions, the predominance of the in-phase Q$^+_2$ mode favors a C-type ordering (C$_{\rm{OO}}$) of the V$^{3+}$ occupied orbitals, in agreement with G$_{\rm{SO}}$ magnetic ground state in the $\pm$1\% region [see Fig.~\ref{Fig5}(b)]. Under large tensile biaxial strain, as shown in Fig.~\ref{Fig5}(b), the out-of-phase Q$^-_2$ JT mode promotes a G-type orbital order (G$_{\rm{OO}}$), which aligns with the C$_{\rm{SO}}$ magnetic ground state. However, we observe that C$_{\rm{SO}}$ exists when Q$^+_2$ dominates i.e. when the amplitude of Q$^-_2$ is zero (below 4\%), whereas C$_{\rm{SO}}$ should be coupled to G$_{\rm{OO}}$ that is favored by Q$^-_2$. Meanwhile, the M$^+_1$ distortion mode increases in the C$_{\rm{SO}}$ region of the phase diagram, raising the question of its role for stabilizing C$_{\rm{SO}}$, substituting Q$^-_2$ as the tensile and compressive biaxial strain increases. For the X$_5^-$ antipolar distortion mode, we observe that its amplitude decreases as the compressive strain decreases and remains rather constant for tensile strain. In the framework of artificially designed RVO$_3$/R'VO$_3$ superlattices, where an hybrid improper ferroelectric behavior~\cite{Bousquet2008,Benedek2011} has been predicted~\cite{Varignon2015}, gaining a deeper understanding of the effect of strain engineering on this distortion could be valuable for tuning the electric polarization of the superlattices. 

Now, we force the out-of-plane to be orthogonal [see Fig.~\ref{Fig5}(d)-(f)].  In Fig.~\ref{Fig5}(d), we observe that the $a^-a^-c^0$ mode's amplitude increases with the substrate lattice parameter though reaching smaller values than in Fig.~\ref{Fig5}(a). On the contrary, the $a^0a^0c^+$ mode remains blocked in both compressively and tensely strained regions (up to 3\%), and above this strain, its amplitude decreases. For tensile strain in Fig.~\ref{Fig5}(e), Q$^+_2$ dominates when the magnetic ground state is G$_{\rm{SO}}$ and Q$^-_2$ dominates when the magnetic ground state is C$_{\rm{SO}}$, whose respective orbital orders are consistently C$_{\rm{OO}}$ and G$_{\rm{OO}}$. For compressive strain, the magnetic ground state is C$_{\rm{SO}}$ despite the absence of Q$^-_2$, raising again the question of the M$^+_1$ distortion mode even against Q$^+_2$. This is likely originating from a strong coupling between strain and JT distortion \cite{Schmitt2020} as the M$^+_1$ is absent in bulk RVO$_3$. In the case of the X$^-_5$ antipolar distortion mode in Fig.~\ref{Fig5}(f), we note that overall its amplitude is diminished compared to the first calculated case in Fig.~\ref{Fig5}(c) and that the amplitude now increases with the biaxial tensile strain. In the light of the calculations made, biaxial strain has strong effect on the GVO film structure and acts on the magnetic phase diagram. As for other bulk orthorhombic perovskite compounds~\cite{Zhou2008,Zhou2010}, lattice distortions and especially cooperative JT distortion are the tuning parameters. The occurrence of a M$^+_1$ distortion mode is also an interesting observation in GVO as it is responsible for the increase of the N\'eel temperature in PrVO$_3$ thin films~\cite{Kumar2019a}.

%
%
%
\subsection{Lattice symmetry by raman spectroscopy}
\begin{figure}[tp]
    \begin{center}
        \includegraphics[width=0.5\textwidth]{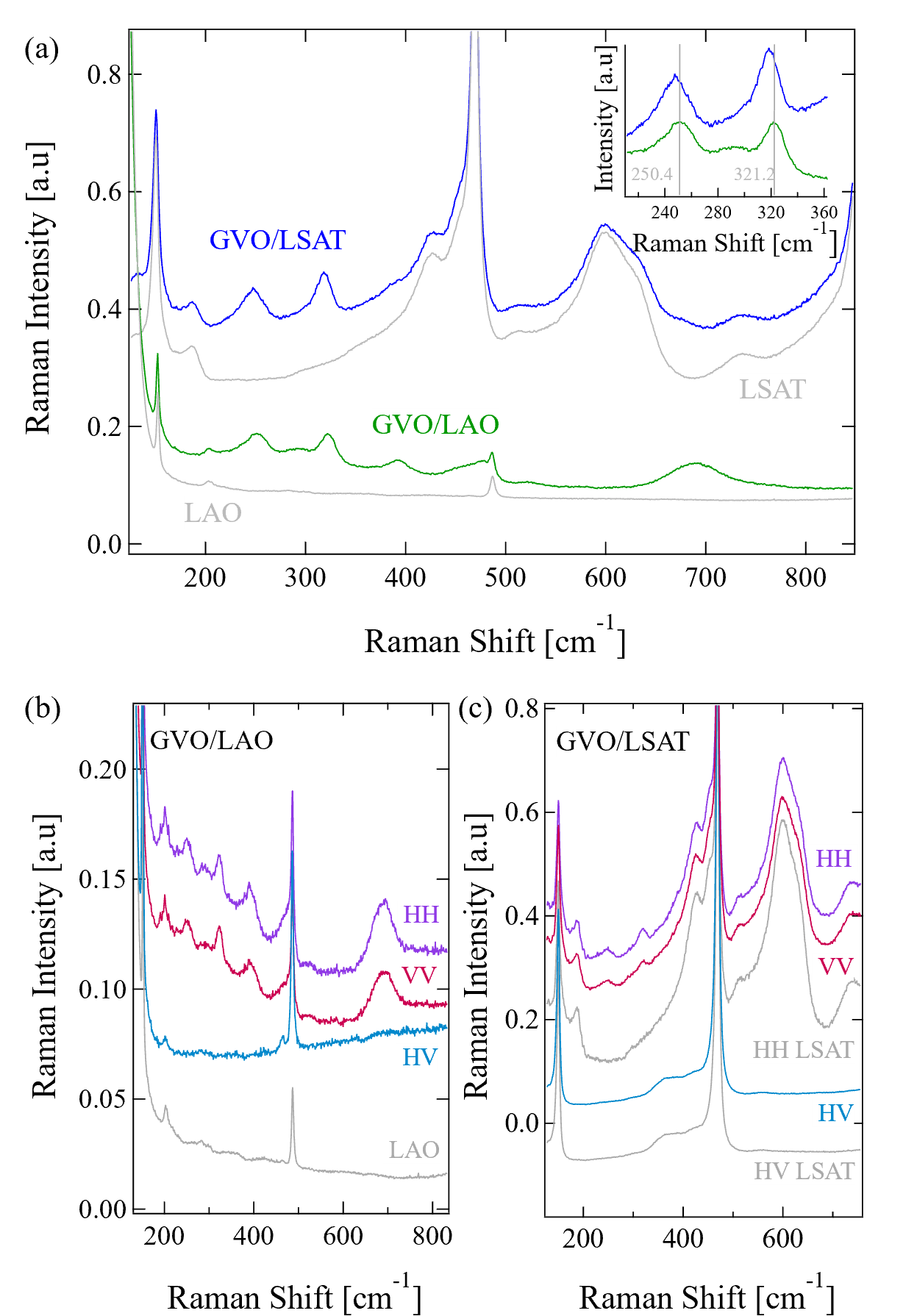}
    \end{center}
\caption{(a) Unpolarized Raman scattering spectroscopy of GVO films grown on LAO and LSAT at room temperature and the bare LAO and LSAT substrate. (b) Polarized Raman scattering spectroscopy of a GVO film grown on LAO at room temperature. H and V stand for horizontal and vertical polarization, respectively and an unpolarized Raman scattering spectroscopy of a bare LAO substrate. (c) Polarized Raman scattering spectroscopy of a GVO film grown on LSAT at room temperature and a bare LSAT substrate.}
\label{Fig6} 
\end{figure}

The orthorhombic $Pbnm$ symmetry allows 24 Raman active modes, 7$A_g$ + 7$B_{1g}$ + 5$B_{2g}$ + 5$B_{3g}$~\cite{Miyasaka2006}. When the lattice undergoes a structural transition to the monoclinic $P2_1/b$ symmetry, $A_g$ and $B_{1g}$ become $A_g$ and $B_{2g}$ and $B_{3g}$ become $B_{g}$ ~\cite{Roberge2015}. In Fig.~\ref{Fig6}(a), we show the unpolarized Raman spectra of GVO grown on LAO and LSAT, together with the spectra of the bare substrates. The other substrates have a much stronger Raman signal, which prevented us from measuring the signal from the films. In Fig.~\ref{Fig6}(b) and (c), spectra were recorded with polarized conditions, whereby the vertical (V) and horizontal (H) polarization directions are parallel to the substrate edges, which correspond to either the $[001]_o$ or the $[110]_o$ ($[\bar 110]_o$) direction of the orthorhombic film.
We start by discussing the unpolarized spectrum of GVO on LAO in Fig.~\ref{Fig6}(a), we observe several Raman bands around 252 cm$^{-1}$, 322 cm$^{-1}$, 393 cm$^{-1}$, 476 cm$^{-1}$, 527 cm$^{-1}$ and 690 cm$^{-1}$ in the unpolarized spectra. According to the literature ~\cite{Miyasaka2006,Miyasaka2003,Miyasaka2005,Roberge2012, Roberge2015,Vrejoiu2016}, the first and the second bands are characteristic of the V-O bond bending and oxygen octahedra rotation modes, respectively. The assignment of the bands at 393 cm$^{-1}$ to lattice-coupled magnetic/orbital excitations, or conversely the lattice vibrations linked to the onset of orbital and magnetic orderings remains unclear ~\cite{Miyasaka2006,Miyasaka2005,Roberge2012, Roberge2015}. The bands at 477 cm$^{-1}$ and 527 cm$^{-1}$ are both attributed to the JT distortion modes. At 690 cm$^{-1}$, another peak exists that is characteristic of a V-O bond stretching mode, which has the $B_{1g}$ symmetry for the orthorhombic $Pbnm$ lattice symmetry and a $A_{g}$ symmetry for the monoclinic $P2_1/b$ lattice. This peak may be convoluted to a phonon density of state, associated to the orbital order in the monoclinic phase ~\cite{Roberge2012, Roberge2015}.
Additionally, in Fig.~\ref{Fig6}(a), we show the unpolarized Raman spectra of GVO deposited on LSAT. On this substrate, fewer GVO peaks are visible due to an overall stronger substrate background. The V-O bond bending and oxygen octahedra rotation modes at 248 cm$^{-1}$ and 318 cm$^{-1}$ are the most prominent. These peaks are shifted by 3 cm$^{-1}$ as compared to the corresponding modes in compressively strained GVO deposited on LAO [see the inset of Fig.~\ref{Fig6}(a)], of the order of 1\%. As epitaxial strain affects the atomic positions (bond angles and lengths), it induces a systematic shift of the corresponding modes, as observed in RVO$_3$ for different R cations~\cite{Miyasaka2006,Miyasaka2003,Miyasaka2005,Roberge2012,Roberge2015}. The peak at 393 cm$^{-1}$ can barely be seen. More importantly, no peak is visible around 700 cm$^{-1}$. 
Now, we show polarized spectra of GVO on LAO and LSAT in Fig.~\ref{Fig6}(b) and (c), respectively. The phonon modes are unambiguously observed for the VV and HH polarization conditions, but the peaks are barely visible for the crossed HV polarization condition. This points to the $A_{g}$ symmetry of the observed peaks in the unpolarized spectra, especially the peak at 690 cm$^{-1}$ for GVO on LAO that is only observed in the monoclinic symmetry ~\cite{Miyasaka2006,Miyasaka2003,Miyasaka2005}. Hence, it indicates a symmetry lowering of the structure of compressively strained GVO at room temperature, whereas the absence of this peak in GVO/LSAT suggests that tensely strained GVO remains orthorhombic when grown on LSAT. This is consistent with the DFT calculations showing a compressive strained favors a $[110]_o$ growth direction, whereas tensile strained favors a $[001]_o$ growth direction, corresponding to a monoclinic $P2_1/m$ and orthorhombic $Pbnm$ symmetry, respectively.

%
%
%
%
\subsection{Influence of biaxial strain on the magnetic properties}
\begin{figure}[tp]
    \begin{center}
        \includegraphics[width=0.5\textwidth]{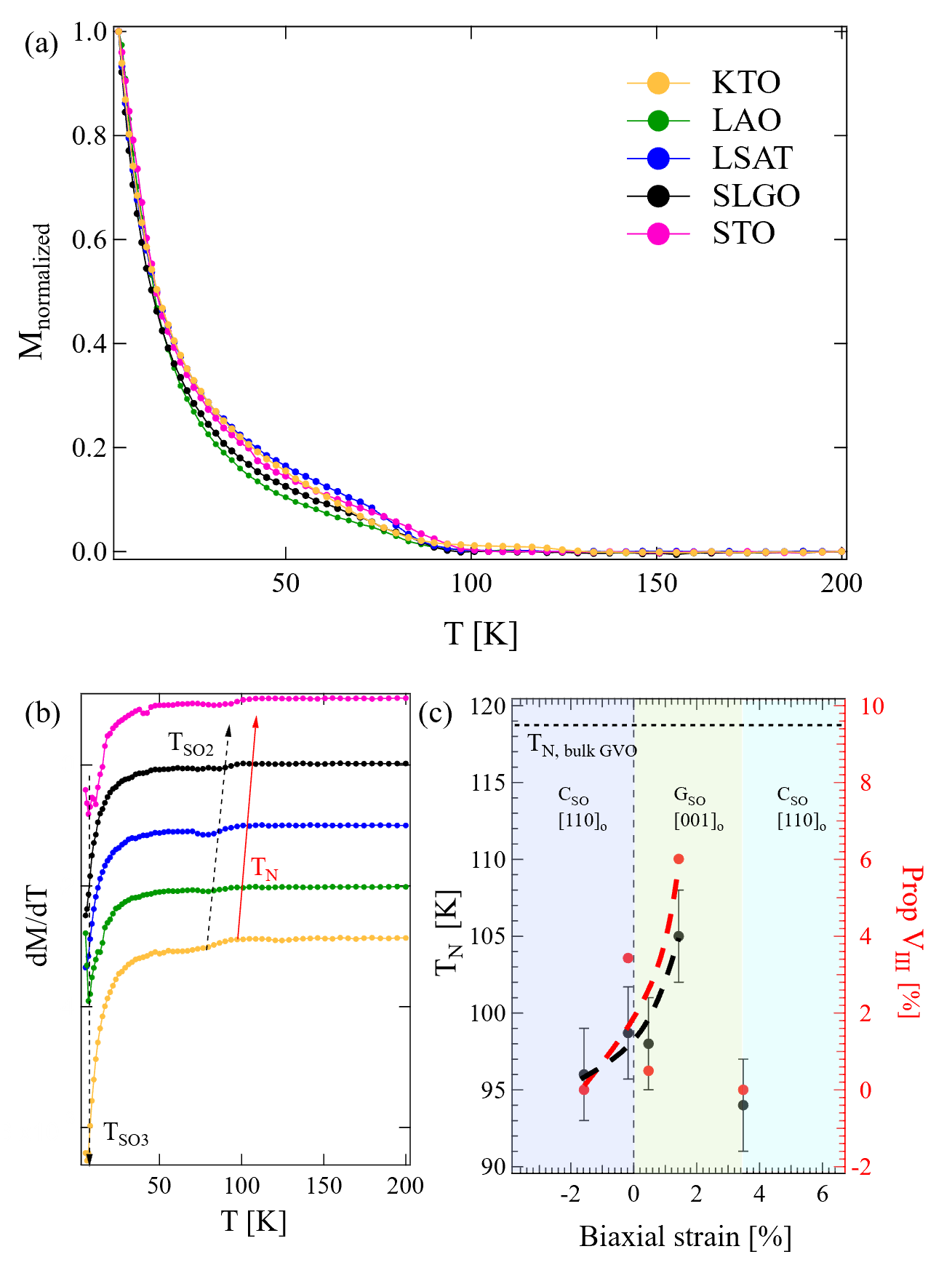}
    \end{center}
\caption{(a) Normalized magnetization as a function of temperature after 5000~Oe field cooling from room temperature, for GVO thin films grown on various substrates. The measurements were performed with an in-plane applied magnetic field of 50~Oe. (b) Corresponding derivative of the magnetic curves, arranged from the lowest T$_{\rm N}$ to the highest T$_{\rm N}$, from bottom to top, respectively. (c) T$_{\rm N}$ and V$_{\rm{III}}$ proportion as a function of the biaxial strain with the results of the symmetry mode analysis overlaid. The last data point for a strain of $+3.48\%$ corresponds to KTO, which is assumed to be relaxed. The dashed lines serve as visual guides}
\label{Fig7} 
\end{figure}

In Fig.~\ref{Fig7}(a), we show the temperature dependence of the magnetization for GVO grown on different substrates. The measurements were carried out with an in-plane applied magnetic field of 50 Oe, after cooling down from room temperature with a field of 5000~Oe. The diamagnetic substrate contribution was corrected to the raw magnetization curve, which was then normalized to the magnetization at 10~K. We clearly see a magnetic signal rising around 100~K, which we attribute to the spin ordering in GVO in agreement with observations in the bulk material~\cite{Tung2006}. The observed magnetic transition corresponds to the antiferromagnetic ordering of the V$^{3+}$ spins at T$_{\rm N}$. The derivatives of the magnetization with respect to temperature are shown in Fig.~\ref{Fig7}(b) and confirm the magnetic transition around 100~K. We observe that T$_{\rm N}$ is modified by the epitaxial strain experienced by the film. Fig.~\ref{Fig7}(c) displays the dependence of T$_{\rm N}$ as a function of the biaxial strain, showing an increase in T$_{\rm N}$ from compressive to tensile strain. The last point, corresponding to KTO, has been marked with an empty circle to represent graphically the assumed relaxation of the film. This increase of the magnetic transition temperature highlights the strengthening of the orbitals overlaps and electron hopping that will enhances the exchange integrals and therefore stabilize the superexchange interactions. On the right side of the figure, we have displayed the proportion of V$_{\rm{III}}$ as a function of biaxial strain, which follows the same trend as T$_{\rm{N}}$. In the background of the figure, the phase diagram deduced from the DFT calculations with the out-of-plane direction kept orthogonal to the substrate plane, showing the possible occurrence of V$_{\rm{III}}$, is superimposed for comparison. These results were selected as they capture the presence of multiple domains. We observe that the proportion of V$_{\rm{III}}$ increases under tensile strain, which is consistent with the results of the DFT calculations that predict a [001]$_o$ orientation for this type of strain. The ground state is C$_{\rm{SO}}$ for tensile strains and G$_{\rm{SO}}$ for compressive strains. Hence, the increase in T$_{\rm{N}}$ with tensile strain can be the signature of the stabilization of different magnetic orders in the films, whose response to a stress imposed by the substrate is different.
Two other transitions can be observed in Fig.~\ref{Fig7}(b), one at T$_{\text{SO2}}$ between 75 K and 90 K, depending on the substrate, highlighted by a dotted arrow, and another at T$_{\text{SO3}}$ between 5 K and 11 K. As T$_{\text{SO2}}$ follows the same trend as T$_{\rm N}$, it could indicate a change/coexistence of the order of the vanadium spins~\cite{Sage2007} and/or the polarization of Gd$^{3+}$ magnetic moment by the antiferromagnetic vanadium sublattice~\cite{Zhao2016,Sasani2021}. At lower temperatures, a transition occurs at T$_{\text{SO3}}$, which seems to be the same for all GVO films grown on different substrates. This transition also observed in polycrystalline compounds \cite{Bozorth1956} at 7.5 K and in single crystals at 8 K \cite{Tung2007} can be assigned to the ordering of Gd$^{3+}$ as for other rare-earths in RVO$_3$~\cite{Reehuis2006,Reehuis2011} at such a low temperature.
\section{Conclusion}
 
In conclusion, we have successfully grown epitaxial and single-phased GdVO$_3$ thin films on top of different single crystal substrates, imposing different values of biaxial strain. The investigation of the structure and the microstructure unveiled the existence of different orientation of the $[001]_o$ direction of the orthorhombic perovskite. Strain engineering allowed to tune the antiferromagnetic transition temperature, which increases with the tensile strain, in contrast with PrVO$_3$ thin films~\cite{Kumar2019a} where compressive strain was decisive for enhancing T$_{\rm N}$. The proportion of domains with out-of-plane orthorhombic long-axis growth follows the same trend as the magnetic transition temperature. DFT calculations revealed that epitaxial strain engineering has a significant influence on structural properties and the magnetic ground state, enabling the tailoring of lattice distortions. The calculations, considering different out-of-plane direction conditions, also highlighted the subtle, and therefore the complex, growth mechanisms that could lead to structural and magnetic phases coexistence. 
Stabilizing only one of these competing phases, \textit{i.e.} a single variant, may be achieved by using orthorhombic instead of cubic substrates as shown recently for YVO$_3$~\cite{Radhakrishnan2024} and LaVO$_3$~\cite{Alexander2024}, involving specific octahedral rotation and cation displacements.

\begin{acknowledgments}
\section{Acknowledgments}
This work was supported by the French National Research Agency (ANR) under project CITRON (ANR-21-CE09-0032), the Région Grand Est under project RHUM (AAP-013-075), and the France 2030 government investment plan under grant PEPR SPIN - SPINMAT (ANR-22-EXSP-0007). The work has benefited of the resources of experimental platforms at IJL: Tube Davm (funded by FEDER EU, ANR, Région Grand Est, and Métropole Grand Nancy), 3M, Magnetism \& Cryogenics and XGamma, supported by the LUE-N4S project, part of the French PIA project Lorraine Université d’Excellence (ANR-15IDEX-04-LUE), and by FEDER-FSE Lorraine et Massif Vosges 2014-2020, an EU program. The first-principles calculations were conducted with the High Performance Computing resources of CRIANN through the projects 2020005 and 2007013 and of CINES through the DARI project A0080911453.
\end{acknowledgments}

\end{document}